# Design of 2D $V_6S_nSe_{6-n}Cl_6$ (n=0, 2, 3, 5) with multilayer kagome lattice and ultrahigh electron mobility

Xing-Yu Wang,[1,2] En-Qi Bao,[1] Su-Yang Shen,[1] Jun-Hui Yuan,[1,*] Jiafu Wang,[1,2,*]

[1]School of Physics and Mechanics, Wuhan University of Technology, Wuhan 430070, China

[2]School of Materials and Microelectronics, Wuhan University of Technology, Wuhan 430070, China

[*]**Corresponding Author**

E-mail: yuanjh90@163.com (J.-H. Yuan); jasper@whut.edu.cn (J. Wang)

**ABSTRACT**

Two-dimensional (2D) kagome materials have attracted considerable attention due to their unique electronic properties. Based on first-principles calculations and employing the "1+3" design strategy, we designed a class of composition-tunable 2D multilayer kagome materials, $V_6S_nSe_{6-n}Cl_6$, and identified four stable structures: $V_6Se_6Cl_6$, $V_6S_2Se_4Cl_6$, $V_6S_3Se_3Cl_6$, and $V_6S_5Se_1Cl_6$. 2D $V_6S_nSe_{6-n}Cl_6$ possesses three kagome layers, two of which are vanadium-based kagome layers, and the other is a sulfur or selenium atomic layer. Electronic structure analysis reveals that 2D $V_6S_nSe_{6-n}Cl_6$ is a narrow direct-bandgap semiconductor with a bandgap ranging from 0.568 to 0.742 eV, and exhibits ultrahigh electron mobility up to $4\times10^4$ cm$^2$ V$^{-1}$ s$^{-1}$. Orbital analysis further demonstrates that the bands contributed by the V-based kagome layers form flat bands and Dirac cones below the Fermi level, and show a relatively high Fermi velocity. In summary, 2D $V_6S_nSe_{6-n}Cl_6$ provides an excellent platform for kagome physics research and the fabrication of nanoelectronic devices, adaptable to various device scenarios.

**Keywords:** Kagome lattice; Two-dimensional materials; Semiconductor; First-principles calculations

# 1. Introduction

At the cutting-edge intersection of condensed matter physics and advanced materials science, the exploration of material systems with unique electronic structures and novel physical properties has always been the core driving force propelling disciplinary progress and technological innovation. In recent years, two-dimensional (2D) materials have become a focal point of attention in both the scientific research community and the industrial sector, owing to their distinctive layered architectures, exceptional physicochemical properties, and immense application potential in numerous fields such as nanoelectronics, optoelectronics, and spintronics[1–4]. Since the discovery of graphene opened a new chapter in 2D material research[5], a variety of 2D materials, including transition metal dichalcogenides (TMDCs)[6], MXenes[7], MBenes[8], single-element metals and semiconductors[9–11], and so on, have continuously emerged, enriching the 2D material family and providing a solid material foundation for constructing novel functional devices with their diverse properties.

Among the myriad of 2D materials, 2D kagome materials stand out due to their unique crystal structures[12,13]. The kagome structure is a two-dimensional periodic network structure composed of triangles and hexagons shalattice vertices. This special geometric configuration endows 2D kagome materials with a series of unique electronic properties. From a theoretical perspective, based on the tight-binding model and band theory, novel band features such as flat bands and Dirac cones emerge in the electronic band structure of 2D Kagome lattice[14,15]. The presence of flat bands results in the effective mass of electrons approaching infinity, with electrons nearly localized within the flat bands, providing an ideal platform for realizing strong-correlation electron effects such as superconductivity, magnetism, and fractional excitations[16–18]. Meanwhile, Dirac cones indicate that electrons near the Fermi level exhibit linear dispersion relations, with electron behavior resembling that of relativistic Dirac fermions, demonstrating unique quantum transport phenomena such as high carrier mobility and the quantum Hall effect[19–21].

In practical research, 2D kagome materials have exhibited numerous remarkable physical properties. For instance, in certain 2D kagome metal systems, the anomalous quantum Hall effect has been observed, which breaks through the reliance of traditional quantum Hall effects on strong magnetic fields and high-quality 2D electron gases, opening up new avenues for developing novel low-power quantum devices[22,23]. In 2D kagome magnetic materials, complex magnetic ordered states and spin liquid states have been discovered, which are of great significance for gaining a deeper understanding of quantum magnetism and developing novel magnetic storage and spintronic devices[24,25]. Additionally, 2D kagome materials also show potential application value in thermoelectric fields, with their unique band structures facilitating efficient thermoelectric conversion [26–28].

However, despite the rich physical properties and broad application prospects of 2D kagome materials, current research still faces numerous challenges. Firstly, the existing 2D kagome material systems are relatively scarce, and there is an urgent need to develop new material systems to provide more candidates for related research[29–31].

Secondly, precisely tuning the electronic properties of 2D kagome materials to achieve on-demand design of their physical properties is a key challenge that be overcome in current research[32,33]. Thirdly, the electronic properties of 2D kagome materials are highly dependent on factors such as their crystal structures, chemical compositions, and dimensional effects[34,35]. Traditional experimental methods often have limitations in regulating these factors, making it difficult to achieve fine control over the electronic properties of materials. For example, dulattice chemical synthesis processes, it is challenging to precisely control atomic-level compositions and defects; when prepalattice 2D materials, it is difficult to ensure uniform layer numbers and crystal quality. These factors limit the in-depth understanding and effective regulation of the electronic properties of 2D kagome materials. First-principles calculations, as a computational simulation method based on quantum mechanics principles, can accurately describe the electronic structures, chemical bonding, and physical properties of materials at the atomic scale, providing a powerful theoretical tool for understanding and regulating the electronic properties of materials. By employing first-principles calculations, we can systematically investigate the influences of different chemical compositions, crystal structures, and dimensional effects on the electronic properties of 2D kagome materials, reveal their underlying physical mechanisms, and thereby guide experimental synthesis and device design.

In this study, based on the "1+3" multilayer kagome 2D material design strategy proposed by our research group in previous work[36–38], we employed first-principles calculation methods to investigate a class of component-tunable 2D vanadium-based kagome materials, namely $V_6S_nSe_{6-n}Cl_6$. This system forms four stable self-supporting 2D multilayer kagome structures through atomic-level construction of three kagome layers. Although the kagome band characteristics of $V_6S_nSe_{6-n}Cl_6$ are not perfect due to orbital coupling effects, by regulating the ratio of sulfur to selenium in the material and altering its chemical composition, we can achieve regulation of electronic properties such as the band structure and carrier mobility.

## 2. Design Strategy

This study sets the design of vanadium (V)-based bilayer kagome materials as its core objective. Relying on the previously proposed "1+3" multilayer kagome design strategy, it achieves precise tuning of the material's physical properties by adjusting the sulfur (S) to selenium (Se) ratio in the $V_6S_nSe_{6-n}Cl_6$. The prototype structure is illustrated in **Figures 1a** and **1b**. Its core feature is the natural formation of upper and lower kagome lattice by V atoms. More importantly, the chalcogen sites (S/Se), denoted as site 1, site 2, and site 3, constitute rich sublattices. Specifically, site 1 atoms reside within the interlayer space between the V-based kagome lattice. Site 2 atoms are located directly above each V-kagome lattice. Crucially, the atoms at site 3 spontaneously self-organize into a third kagome lattice. This emergent kagome lattice, interwoven with the primary V-based ones, establishes a foundation for introducing strong geometric frustration and diverse electronic orbital interactions.

The focus of this study lies in precisely regulating the aforementioned multiple kagome networks by systematically varying the value of n (*i.e.*, the number of selenium atoms substituted by sulfur atoms). As shown in **Figures 1c-1j**, a series of distinct compounds is generated as n increases from 0 to 6. Notably, at n=3, two different compounds are possible depending on the specific sites occupied by S/Se: $V_6S_3Se_3Cl_6$ (S occupies site 1 and site 2, while Se occupies site 3) and $V_6Se_3S_3Cl_6$ (S occupies site 3, while Se occupies site 1 and site 2). The potentially stable compounds resulting from this S/Se ratio tuning include $V_6Se_6Cl_6$, $V_6Se_5S_1Cl_6$, $V_6Se_4S_2Cl_6$, $V_6Se_3S_3Cl_6$, $V_6S_3Se_3Cl_6$, $V_6S_4Se_2Cl_6$, $V_6S_5Se_1Cl_6$ and $V_6S_6Cl_6$.

Subsequent calculations based on this design strategy were performed for the V-based multilayer kagome materials using the Vienna *Ab initio* Simulation Package (VASP).[39,40] The detailed computational settings refer to **Note 1** in the **Supplementary Materials**.

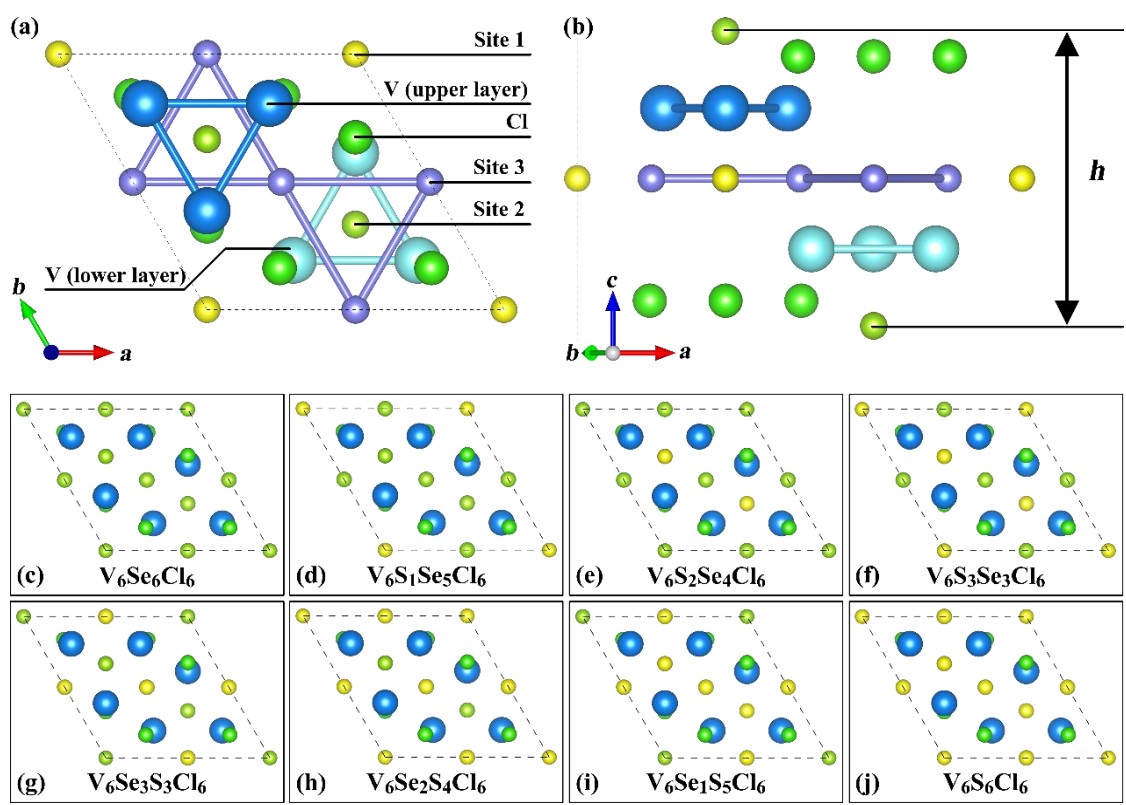

**Figure 1.** Schematic illustrations of the design concept for V-based multilayer kagome materials. (a) Top and (b) side view of the $V_6S_nSe_{6-n}Cl_6$ monolayer, showing the positions of V atoms in the upper and lower layers as well as sites such as Cl atoms.; (c-j) Top view of the $V_6S_nSe_{6-n}Cl_6$ monolayers with n varying from 0 to 6. In (f), at n = 3, S atoms occupy site 1 and site 2; In (g), also at n = 3, S atoms occupy site 3. Different atomic species are represented by colored spheres for clarity.

## 3. Results and discussions

To ensure the reliability of subsequent property investigations, we first conducted a systematic screening of the structural stability within the designed $V_6S_nSe_{6-n}Cl_6$ material family. Phonon dispersion calculations are crucial for assessing kinetic stability. As shown in **Figure 2a**, the phonon spectrum of $V_6Se_6Cl_6$ (n=0) shows no imaginary frequencies in the entire Brillouin zone. In addition, as presented in **Figures S1-S3** of the **Supplementary Material**, the phonon spectra of $V_6S_2Se_4Cl_6$ (n=2), $V_6Se_3S_3Cl_6$ (n=3), and $V_6Se_1S_3Cl_6$ (n=5) also exhibit no imaginary frequencies (the tiny imaginary frequency at the Γ point is attributed to computational errors). This indicates that these structures reside at local minima on the potential energy surface, with stable lattice vibration modes, thereby satisfying the criterion for kinetic stability. It is noteworthy that for n=3, there are two isomers. We denote the structure with S atoms occupying site1 and site 2, and Se atoms occupying site 3, as $V_6S_3Se_3Cl_6$. The structure with Se atoms occupying site 1 and site 2, and S atoms occupying site 3, is denoted as $V_6Se_3S_3Cl_6$. Phonon dispersion calculations show that the stable structure is $V_6Se_3S_3Cl_6$. Therefore, the term "n=3" subsequently refers to the $V_6Se_3S_3Cl_6$ monolayer.

To analyze the imaginary-frequency modes of unstable materials, the phonon vibration modes of $V_6S_1Se_5Cl_6$, $V_6S_3Se_3Cl_6$, $V_6Se_2S_4Cl_6$, and $V_6S_6Cl_6$ were calculated. Taking $V_6S_6Cl_6$ as an example, its results are shown in **Figures 2b-2d** (other results show in **Figures S4-S6**). **Figure 2b** shows the phonon dispersion, where the presence of imaginary-frequency branches near the Γ point confirms the kinetic instability of the $V_6S_6Cl_6$ monolayer. **Figure 2c-d** illustrates the optical branch vibration mode with an imaginary frequency of -1.44 THz at the Γ point. The results indicate that this imaginary frequency mode primarily arises from the relative translation between the V atomic layer and the S atomic layer. Specifically, the outermost S atoms (site 2) vibrate in the opposite direction to the central-layer S atoms, while the V atomic layer exhibits a tendency to separate along the *c*-direction. Such a large imaginary frequency suggests that $V_6S_6Cl_6$ requires significant lattice distortion or lattice reconstruction along the vibration direction to reach a minimum point on the potential energy surface, thereby achieving a possible stable structure. The imaginary frequency modes of the other three unstable structures are similar to those of $V_6S_6Cl_6$, with the results presented in **Figures S4-S6**.

To further verify the thermal stability of $V_6S_nSe_{6-n}Cl_6$ (n = 0, 2, 3, 5) monolayer at finite temperatures, AIMD simulations were performed. **Figure S7** show the final snapshots of the structures after evolution at high temperatures of 900 K or 1200 K. The simulation results demonstrate that the final crystal structures of $V_6S_2Se_4Cl_6$, $V_6Se_3S_3Cl_6$, and $V_6Se_1S_5Cl_6$ remained intact without bond breaking or structural collapse at 1200 K. Similarly, $V_6Se_6Cl_6$ maintained its structural integrity at 900 K. The inset figures provide a clear structural comparison before and after AIMD simulations, offering robust evidence for the compounds' exceptional thermal stability and their capacity to maintain structural integrity at elevated temperatures.

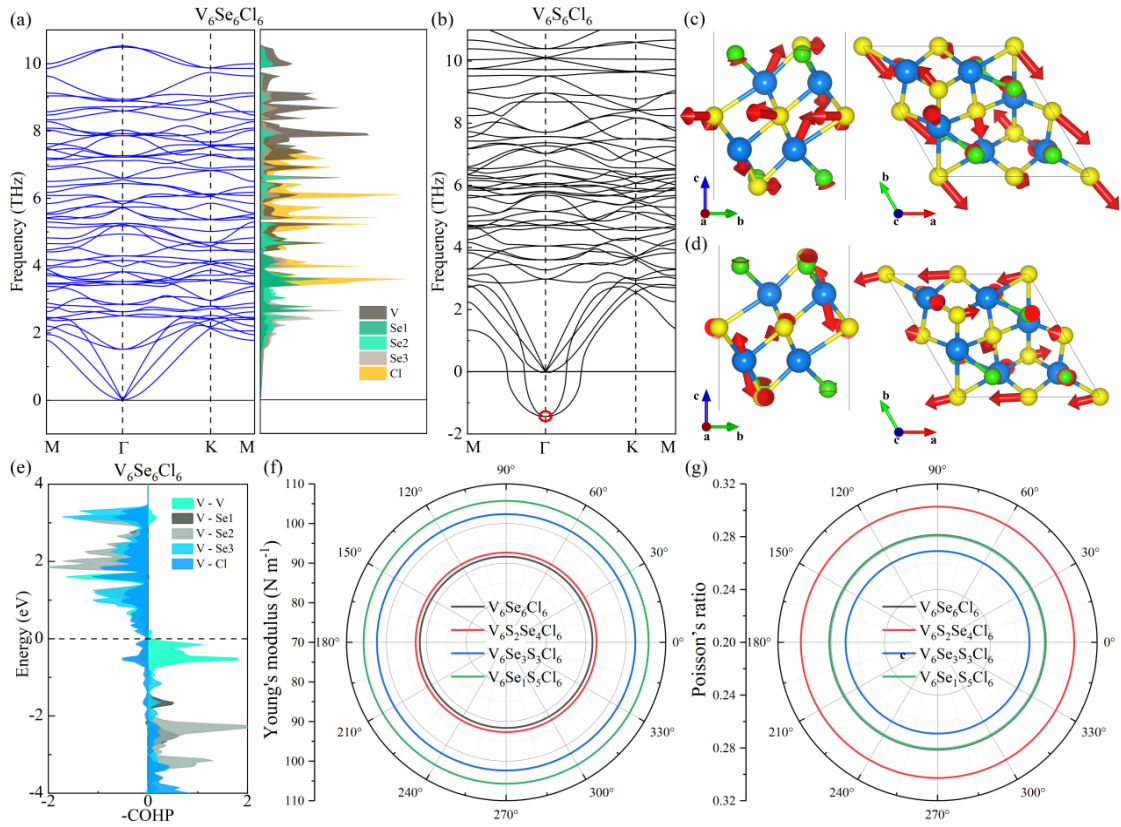

**Figure 2.** (a) Phonon dispersion and phonon density of states of $V_6Se_6Cl_6$ monolayer, (b) phonon dispersion of the $V_6S_6Cl_6$ monolayer, side and top views of the two phonon vibration modes of $V_6S_6Cl_6$ monolayer (c-d) with the frequency of -1.44 THz around the $\Gamma$ point, the length of the arrows represents the amplitude of the atomic vibrations, (e) the crystal orbital Hamilton population analysis of $V_6Se_6Cl_6$ monolayer, the calculated angle-dependent in-plane (f) Young's modulus and (g) Poisson's ratio of $V_6S_nSe_{6-n}Cl_6$ (n = 0, 2, 3, 5) monolayer.

To further verify the stability origin of the four screened $V_6S_nSe_{6-n}Cl_6$ monolayers, we systematically analyzed their bonding characteristics. Bader charge calculations[41] and crystal orbital Hamilton population (COHP) analysis[42,43] were performed. These methods reveal the influence of composition tuning on stability through charge transfer and chemical bond strength. Bader charge analysis results are listed in **Table S1**. They show a clear trend of electron transfer from V atoms to the surrounding nonmetal atoms (S/Se/Cl). This suggests significant ionic character in the bonding. In $V_6Se_6Cl_6$ (n=0), each V atom loses approximately 1.243 electrons. As the S content increases, the charge loss per V atom systematically rises: 1.287 (n=2), 1.301 (n=3), and 1.352 (n=5). This trend indicates that introducing S enhances the electronegativity difference between V and the nonmetal atoms. Consequently, stronger ionic bonding components form.

COHP analysis provides key evidence from the perspective of bonding/antibonding interactions. The results show that bonding states dominate below the Fermi level for both V-V and V-S (site 2) bonds. Furthermore, the occupancy of antibonding states is very low, as shown in **Figure 2e** and **Figure S8**. This favorable bonding characteristic aligns well with our design concept. The strong bonding formed by passivation with nonmetal elements effectively maintains the stability of the kagome lattice.

We further evaluated the mechanical stability of these four monolayers. According to the Born-Huang criteria[44], a 2D material is mechanically stable if it satisfies $C_{11}C_{22} - C_{12}^2 > 0$ and $C_{66} > 0$. The calculated elastic constants for $V_6S_nSe_{6-n}Cl_6$ are presented in **Table S2**. Evidently, all four predicted monolayers meet these mechanical stability criteria. This confirms that the four V-based kagome monolayers predicted in this work can theoretically exist as free-standing 2D materials, laying a theoretical foundation for relevant experimental exploration.

**Table 1.** Calculated lattice constant $a/b$ (Å), bond length $l$ (Å), band gaps (eV) of $V_6S_nSe_{6-n}Cl_6$ (n = 0, 2, 3, 5) monolayers.

| Materials | $a/b$ | $h$ | $l_{V-V}$ | $l_{V-Site\ 1}$ | $l_{V-Site\ 2}$ | $l_{V-Site\ 3}$ | $l_{V-Cl}$ | $E_g^{PBE}$ | $E_g^{HSE06}$ |
|---|---|---|---|---|---|---|---|---|---|
| $V_6Se_6Cl_6$ | 6.770 | 5.617 | 2.841 | 2.786 | 2.413 | 2.433 | 2.425 | 0 | 0.742 |
| $V_6S_2Se_4Cl_6$ | 6.733 | 5.654 | 2.803 | 2.794 | 2.275 | 2.431 | 2.427 | 0 | 0.740 |
| $V_6Se_3S_3Cl_6$ | 6.685 | 5.423 | 2.775 | 2.695 | 2.409 | 2.317 | 2.436 | 0 | 0.568 |
| $V_6Se_1S_5Cl_6$ | 6.652 | 5.455 | 2.744 | 2.701 | 2.272 | 2.315 | 2.436 | 0 | 0.591 |

**Table 2** summarizes the key structural parameters of the four stable compounds. As the sulfur content increases (n varying from 0 to 5), the lattice constant $a/b$ exhibits a systematic reduction from 6.770 Å to 6.652 Å. This contraction can be attributed to the higher proportion of shorter V-S bonds compared to V-Se bonds. Correspondingly, the V-V bond length decreases from 2.841 Å to 2.744 Å, indicating enhanced in-plane bonding strength upon sulfur incorporation. Furthermore, variations in the V-S(Se) bond lengths at different sites (Site 1~3) highlight the local bonding heterogeneity, providing crucial structural insights into how compositional tuning influences the stability of these materials.

The screening results highlight the significant influence of the S/Se ratio on structural stability. The fact that not all stoichiometric configurations are stable underscores that specific atomic arrangements and component matching are crucial for achieving the lowest-energy stable structure in the $V_6S_nSe_{6-n}Cl_6$ system. The successful identification of these stable phases ensures that subsequent investigations into mechanical properties, bonding analysis, electronic structure, and carrier mobility are conducted on thermodynamically viable structures, greatly enhancing the reliability and physical significance of the conclusions drawn in this study.

The mechanical properties of materials, particularly stiffness and toughness, are critical indicators for assessing structural robustness and potential for device integration. Based on the elastic constants $C_{ij}$, we calculated the in-plane Young's modulus $Y$ and Poisson's ratio $v$ of the $V_6S_nSe_{6-n}Cl_6$ (n = 0, 2, 3, 5) monolayers. The method used in the calculation refer to **Note 2** in the **Supplementary Material**. Our study reveals that the $V_6S_nSe_{6-n}Cl_6$ (n = 0, 2, 3, 5) monolayers exhibit remarkable isotropic mechanical behavior in the in-plane direction, as shown in **Figures 2f-2g**. This characteristic is highly advantageous for applications in flexible electronics and nano-electromechanical systems (NEMS), where a uniform mechanical response ensures performance stability and reliability under complex stress conditions.

Numerically, $Y$ of these four materials range from 91.62 N m$^{-1}$ to 105.7 N m$^{-1}$, indicating a moderate in-plane stiffness that is comparable to many other 2D semiconductors, such as $Si_3N$ (106.3 N m$^{-1}$)[45], $Sc_6S_6Cl_6$ (70.74 N m$^{-1}$)[37]. Concurrently, their positive

Poisson's ratios ($v$ ranging from 0.269 to 0.303) suggest standard ductile behavior. A key finding of this research is the systematic stiffening effect induced by sulfur substitution. According to the data in **Table S2**, the Young's modulus shows a clear increasing trend with higher sulfur content (increasing n value). This phenomenon can be attributed to the shorter and stronger V-S bonds compared to V-Se bonds, which enhance the overall rigidity of the kagome lattice. This tunability demonstrates that compositional engineered lattice via the S/Se ratio not only provides an effective means for electronic structure modulation but also enables precise adjustment of the mechanical performance of this material family, allowing for optimization of mechanical properties according to specific application requirements.

To gain deeper insight into the electronic structure of $V_6S_nSe_{6-n}Cl_6$ monolayers, particularly features associated with the kagome lattice, such as flat bands and Dirac cones, we systematically computed and analyzed their band structures. The results are presented in **Figures 3** and **S9**. GGA-PBE calculations (orange dashed lines in **Figure 3**) reveal that all $V_6S_nSe_{6-n}Cl_6$ (n = 0, 2, 3, 5) monolayers exhibit semi-metallic character along the Γ–Γ path. However, the accuracy of GGA-PBE in predicting band gaps for semiconductors and insulators is often limited[46–49]. Therefore, we re-evaluated the band structures using the HSE06 hybrid functional[50]. As shown in **Figures 3a-3d** (black solid lines), the HSE06-calculated band gaps are significantly larger, measuring 0.742 eV for $V_6Se_6Cl_6$, 0.740 eV for $V_6S_2Se_4Cl_6$, 0.568 eV for $V_6Se_3S_3Cl_6$, and 0.591 eV for $V_6Se_1S_5Cl_6$, respectively. Moreover, they are all direct-bandgap semiconductors at the Γ–Γ point. Consequently, since the bandgap obtained from GGA-PBE calculations is zero, the subsequent carrier mobility calculations were all performed using the results from the HSE06 functional.

To further elucidate the orbital contributions to the electronic structure, we analyzed the orbital-projected band structures. Based on differences in the *p*-orbital contributions from S, Se, and Cl atoms, the systems can be broadly categorized into two types. We present only the results for $V_6Se_6Cl_6$ (**Figures 3e-3j**) in the main text. The corresponding results for $V_6S_2Se_4Cl_6$, $V_6Se_3S_3Cl_6$ and $V_6Se_1S_5Cl_6$ are provided in **Figures S10-S12**. The results indicate that the conduction band and the top of the valence band (within –1 eV to 2 eV) in all four materials are primarily dominated by V-*d* orbitals. Specifically, the conduction band is mainly contributed by the V-*dz²* orbital, with additional involvement from the V-*dx²-y²* orbital. The valence band near the Fermi level is also composed of V-*dz²* and V-*dx²-y²* orbitals. Notable differences appear at deeper energy levels (approximately -2 eV to -1 eV below the Fermi level). For $V_6Se_6Cl_6$ and $V_6S_2Se_4Cl_6$, the primary contributions in this energy range come from Site1-*pz*, Site3-*px*, Site3-*py*, and Cl-*pz* orbitals. In contrast, for $V_6Se_3S_3Cl_6$ and $V_6Se_1S_5Cl_6$, the contribution from Site3 orbitals decreases, while those from Cl-*px* and Cl-*py* orbitals increase. Overall, the dominant contributions in this region come from Site1-*p* and Cl-*p* orbitals.

This orbital-resolved analysis confirms that all four materials belong to a V-*d*-orbital-dominated kagome system, with *p*-orbital contributions serving as a secondary tuning parameter. The distinct *p*-orbital participation patterns between the two categories (n =

0, 2 vs. n = 3, 5) highlight the effect of sulfur substitution on modifying the orbital hybridization and electronic environment in this tunable kagome material family.

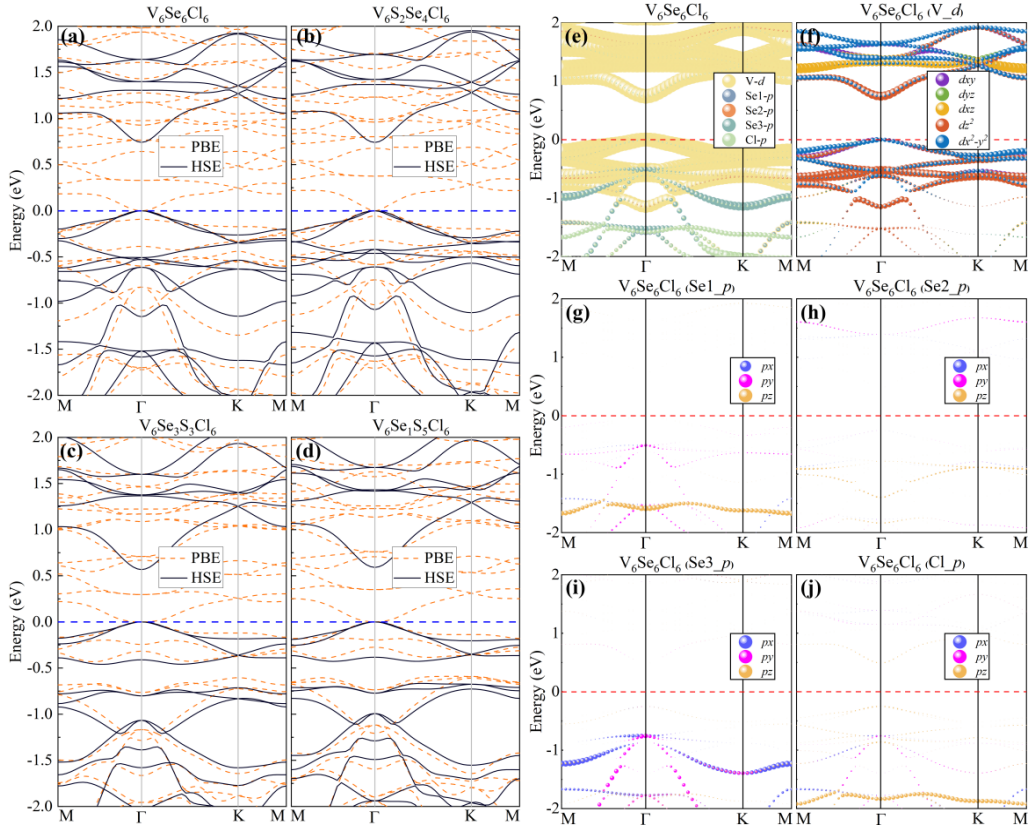

**Figure 3.** (a-d) Calculated band structures of $V_6S_nSe_{6-n}Cl_6$ (n = 0, 2, 3, 5) monolayers with the HSE06 hybrid functional and GGA-PBE computational methods. (e-j) Projected band structure of $V_6Se_6Cl_6$ monolayer based on the HSE06 hybrid functional.

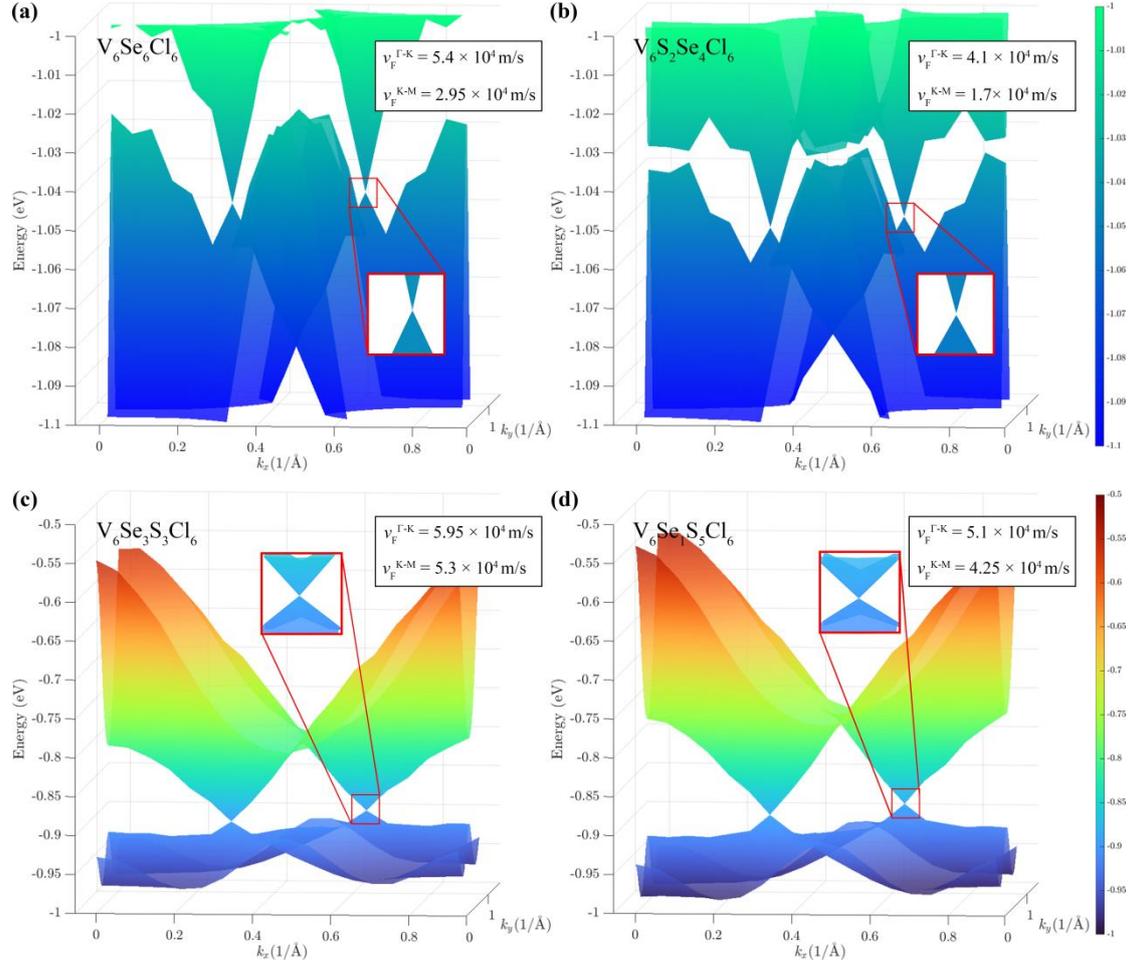

**Figure 4.** 3D band structures of $V_6S_nSe_{6-n}Cl_6$ (n = 0, 2, 3, 5) monolayers based on HSE06 calculations. (a) $V_6Se_6Cl_6$, (b) $V_6S_2Se_4Cl_6$, (c) $V_6Se_3S_3Cl_6$, (d) $V_6Se_1S_5Cl_6$.

Furthermore, for kagome materials, their characteristic electronic features—such as flat bands and Dirac cones in the electronic band structure—are of particular importance. As proposed in our earlier study on the $Sc_6S_6X_6$ (X = Cl, Br, I) system[37], its kagome lattice derived bands lie in the conduction band, which poses significant challenges for both experimental observation and practical applications. Notably, in the $V_6S_nSe_{6-n}Cl_6$ (n = 0, 2, 3, 5) monolayers presented in this work, the flat bands and Dirac cones contributed by the V-based kagome lattice are located below the Fermi level, which is highly favorable for experimental detection and practical implementation. However, as seen from the band structure, the flat band in 2D $V_6S_nSe_{6-n}Cl_6$ (specifically, the first valence band below the Fermi level) already exhibits pronounced dispersion due to orbital hybridization, which is particularly evident in $V_6Se_6Cl_6$ and $V_6S_2Se_4Cl_6$, as shown in **Figures 3a** and **3b**. In contrast, the situation is relatively better in $V_6S_3Se_3Cl_6$ and $V_6S_5Se_1Cl_6$, where the flat-band character remains well preserved along the G–K–M path. In addition, Dirac cone structures are clearly observed at the K point below the Fermi level in all four materials. **Figure 4** presents the 3D band structures of the 2D $V_6S_nSe_{6-n}Cl_6$ (n = 0, 2, 3, 5). This directly confirms the presence of Dirac cones in this series, forming the electronic structural basis for their unique charge transport properties. The Fermi velocity $v_F$ quantifies the slope of the Dirac cone. In the Γ–K direction, $v_F$ varies significantly from $4.1×10^4$ m/s in $V_6S_2Se_4Cl_6$ to $5.95 × 10^4$ m/s in $V_6Se_3S_3Cl_6$. In the Γ–M direction, the variation is more pronounced, from $1.7 × 10^4$ m/s in $V_6S_2Se_4Cl_6$ to $5.3 × 10^4$ m/s in $V_6Se_3S_3Cl_6$. The Fermi velocity near the Dirac cone at the K point in 2D $V_6S_nSe_{6-n}Cl_6$ is significantly lower than that at the Dirac cone of graphene ($8.22×10^5$ m/s) [51], which is primarily attributed to interlayer orbital coupling among the multiple atomic layers. Additionally, in these 2D kagome materials, varying the S/Se composition acts as a powerful "tuning knob". This provides a crucial theoretical foundation and material design space for tailing specific electronic or spintronic devices.

Although the kagome flat band characteristics in $V_6S_nSe_{6-n}Cl_6$ monolayer are not prominent, it remains an excellent direct band gap semiconductor. For semiconductors, carrier mobility is one of the key factors in assessing their applicability in high-performance devices. Therefore, based on the modified deformation potential theory proposed by Lang *et al.*[52], we evaluated the phonon-limited carrier mobility of $V_6S_nSe_{6-n}Cl_6$ monolayer. The computational details refer to **Note 3** in the **Supplementary Material**. In our calculations, we employed HSE06 to determine the band structure of its corresponding orthorhombic cell. The carrier effective masses and corresponding carrier mobilities of 2D $V_6S_nSe_{6-n}Cl_6$ are shown in **Table 2**.

Analysis of carrier effective mass shows that electron effective masses ($m_e \approx 0.41\sim0.46\ m_0$) are significantly lower than hole effective masses ($m_h \approx 1.08\sim1.55\ m_0$). The relatively large hole effective mass is primarily due to the flat-band character of the kagome lattice in 2D $V_6S_nSe_{6-n}Cl_6$. Furthermore, the deformation potential constants for electron ($|E_{1e}| = 0.41\sim1.52$ eV) are generally smaller than hole deformation constants ($|E_{1h}| = 1.63\sim2.21$ eV). Smaller deformation potentials suggest weaker electron-phonon scatting for electrons. The calculated mobility values show significant performance advantages and clear anisotropy. Particularly, electron mobility ($\mu_e^{2D}$) of 2D $V_6S_nSe_{6-n}Cl_6$ reaches up to $4\times10^4$ cm$^2$V$^{-1}$s$^{-1}$, much higher than hole mobility ($\mu_h^{2D} \approx 290–415$ cm$^2$V$^{-1}$s$^{-1}$).

On the other hand, S/Se composition tuning significantly affects carrier transport properties. As sulfur content increases (n from 0 to 5), the in-plane elastic modulus $C^{2D}$ systematically increases (from ∼100 N m$^{-1}$ to ∼110 N m$^{-1}$). This indicates enhanced lattice stiffness, which helps reduce deformation potential scattering. However, mobility does not change monotonically. $V_6Se_3S_3Cl_6$ (n=3) exhibits the highest electron mobility.

This optimal performance results from the best balance between its smallest deformation potential constant ($|E_{1e,a}|$ = 0.41 eV) and moderate effective mass. These results suggest that moderate S substitution (n=3) optimally optimizes the trade-off between phonon scatting and carrier effective mass, leading to peak mobility performance.

Overall, $V_6S_nSe_{6-n}Cl_6$ monolayers, particularly $V_6Se_3S_3Cl_6$, exhibit high electron mobility comparable to high-performance 2D semiconductors like phosphorene (~ $10^4$ cm$^2$V$^{-1}$s$^{-1}$)[53]. Combined with their tunable band gaps and pronounced anisotropy, these materials show great promise for field-effect transistors and other nanoelectronic devices. Precise control of S/Se ratio provides an effective approach to optimize their carrier transport properties.

**Table 2.** Calculated effective mass ($m^*$, $m_0$) of electron (*e*) and hole (*h*), in-plane elastic modulus $C_{2D}$ (N m$^{-1}$), deformation potential constant ($|E_1|$, eV) and corresponding carrier mobility (μ, cm$^2$ V$^{-1}$ s$^{-1}$) of $V_6S_nSe_{6-n}Cl_6$ (n = 0, 2, 3, 5) monolayers.

| Materials | Carrier | $m_a^*$ | $m_b^*$ | $|E_{1a}|$ | $|E_{1b}|$ | $C_a^{2D}$ | $C_b^{2D}$ | $\mu_a^{2D}$ | $\mu_b^{2D}$ |
|---|---|---|---|---|---|---|---|---|---|
| $V_6Se_6Cl_6$ | e | 0.425 | 0.426 | 1.060 | 1.046 | 100.17 | 99.81 | 10590.01 | 10626.05 |
|  | h | 1.215 | 1.171 | 1.954 | 1.891 |  |  | 394.49 | 415.70 |
| $V_6S_2Se_4Cl_6$ | e | 0.411 | 0.411 | 1.529 | 1.356 | 103.62 | 103.17 | 6080.60 | 6449.12 |
|  | h | 1.315 | 1.084 | 2.213 | 2.167 |  |  | 291.07 | 356.43 |
| $V_6Se_3S_3Cl_6$ | e | 0.458 | 0.458 | 0.414 | 0.671 | 105.38 | 105.70 | 40630.40 | 32157.95 |
|  | h | 1.442 | 1.554 | 1.633 | 1.631 |  |  | 390.70 | 363.04 |
| $V_6Se_1S_5Cl_6$ | e | 0.451 | 0.451 | 0.513 | 0.719 | 110.37 | 110.83 | 33032.66 | 27997.48 |
|  | h | 1.406 | 1.495 | 1.921 | 1.887 |  |  | 317.17 | 301.27 |

## 4. Conclusion

In summary, based on first-principles calculations and the "1+3" design strategy proposed in our earlier work, we successfully predicted a new class of composition-tunable 2D vanadium-based multilayer kagome monolayers, $V_6S_nSe_{6-n}Cl_6$. Stability analysis confirmed four independently stable structures: $V_6Se_6Cl_6$, $V_6S_2Se_4Cl_6$, $V_6S_3Se_3Cl_6$, and $V_6S_5Se_1Cl_6$. Electronic structure analysis shows that 2D $V_6S_nSe_{6-n}Cl_6$ are all narrow direct-bandgap semiconductors with bandgaps ranging from 0.568 to 0.742 eV. Below the Fermi level, the valence bands feature distinctly dispersive flat bands and well-defined Dirac cones contributed by the V-based kagome lattice. The fitted Fermi velocity near the Dirac cones reaches $1.7–5.95×10^4$ m/s. Furthermore, carrier mobility limited by phonon scattering exhibits pronounced electron–hole asymmetry: the electron mobility is as high as $0.61–4.06×10^4$ $cm^2V^{-1}s^{-1}$, while the hole mobility is only 290–415 $cm^2V^{-1}s^{-1}$. The lower hole mobility is primarily attributed to the larger hole effective mass induced by the flat band. The results also indicate that adjusting the S/Se ratio allows tunability of the stability, electronic structure, and mechanical properties of 2D $V_6S_nSe_{6-n}Cl_6$ to a certain extent. Therefore, our work not only expands the research scope of 2D multilayer kagome materials but also provides a theoretical foundation for further investigation of 2D $V_6S_nSe_{6-n}Cl_6$.

## Acknowledgements

This work was supported by the Fundamental Research Funds for the Central Universities (WUT: 2024IVA052), the National Undergraduate Innovation and Entrepreneurship Training Program (Grant No. 202510497080). We would like to express our gratitude for the support from the High-performance Computing Platform of Wuhan University of Technology.

Entrepreneurship Training Program (Grant No. 202510497080). We would like to express our gratitude for the support from the High-performance Computing Platform of Wuhan University of Technology.